\documentclass[intlimits,twoside,a4paper]{article}

\usepackage{caption}
\usepackage{subcaption}
\usepackage{IEEEtrantools}
\usepackage{tabularray}
\usepackage[eqsecnum]{cmpj3}
\articletype{A part of the Special collection to the 100th anniversary of the birth of Ihor Yukhnovskii}
\issue{2025}{28}{2}{23603}
\doinumber{10.5488/CMP.28.23603}

\title[Liquid-gas state regularities viewed from the Ising model perspective]%
{Liquid-gas state regularities as  a manifestation of global isomorphism with the Ising model}

\author[L.~A. Bulavin, V.~L.~Kulinskii, A. M.~Katts, A. M.~Maslechko]{
	L.~A. Bulavin\orcid{0000-0002-8063-6441}\refaddr{knu},
	V.~L.~Kulinskii\orcid{0000-0002-5139-843X}\refaddr{onu,ontu}
	\thanks{Corresponding author: \email{kulinskij@onu.edu.ua}.},
	A. M.~Katts\orcid{0009-0009-3656-9446}\refaddr{onu},
	A. M.~Maslechko\orcid{0000-0002-1965-4499}\refaddr{onu}}
\addresses{\addr{knu} Department of Molecular Physics,  Taras Shevchenko National University of Kyiv, 2 Academician Glushkov Prosp., 03022 Kyiv, Ukraine 
	\addr{onu}  Faculty of Mathematics, Physics and Information Technologies, Odesa I. I. Mechnikov National University, 2~V.~Zmijenka St., 65026 Odesa, Ukraine 
	\addr{ontu} Department of Physics and Mathematics, Odesa National University of Technology, 112 Kanatna St., 65039 Odesa, Ukraine}
\Keywords{liquid-gas phase transition, critical phenomena, Ising model}
\date{Received March 9, 2025, in final form May 13, 2025}

\begin{document}
	\maketitle
	\begin{abstract}
		Liquid-gas  equilibrium is considered using the global isomorphism with the Ising-like (lattice gas) model. Such an approach assumes the  existence of the order parameter in terms of which the symmetry of binodal is restored not only in the vicinity of the critical point (critical isomorphism) but  also globally in the whole coexistence region. We show how the empirical law of the rectilinear density diameter of the liquid-gas binodal allows us to derive a rather simple form of the isomorphism transformation between the fluid and lattice gas model of Ising-type. The relations for critical parameters which follow from such isomorphism are tested on a variety of fluid systems, both real and model ones. Moreover, we consider the phase equilibrium in polymer solutions and the Flory $\theta$-point as the extreme state of such equilibrium within our approach. The most crucial testing in 2D case is using the Onsager exact solution of the Ising model, and we represent the results of our approach to the calculation of critical point parameters of monolayers for noble gases and the surface tension.
		\printkeywords
	\end{abstract}
\section{Introduction}\label{sec_intro}
Symmetry is a unified concept in physics and plays a great role in the study of phase transitions. If the correlation functions of two equilibrium states have the same symmetry, then there is no restriction to the transformation of one phase into another under a continuous change of state parameters \cite{book_ll5_en}. The liquid-gas equilibrium in simple fluids is a characteristic example here. The statement that there is no qualitative difference between liquid and gas is due to van der Waals \cite{vdw_nobelecture1910}. It led him to the seminal equation of state which for the first time predicted a critical point purely analytically. On this basis, the principle of corresponding states (PCS) was formulated. The further development of the theory of critical phenomena has stressed the importance of symmetry, where Ising-like lattice models \cite{crit_onsager_pr1944,book_baxterexact} played a crucial role in establishing the universality classes of criticality \cite{crit_zinn2021quantum}. The origin of universal critical behaviour is the correlation length divergence, which makes the difference between the discrete lattice model and the continuous fluid irrelevant for the leading singular terms of thermodynamic quantities in the vicinity of the critical point. Beyond the asymptotic region, the main difference is caused by the absence of particle-hole symmetry for real and model fluids which leads to the asymmetry of a liquid-gas density binodal. The short-range hard-core part of the interparticle potential is treated differently from the attractive long-range part, as it causes an entropic effect of the excluded volume \cite{crit_hubbardschofield_pla1972, book_yukhgol_en}. The natural characteristic of such an asymmetry between coexisting phases is the density diameter:
	\begin{equation}\label{eq:diam}
		\tilde{\rho}_d = \frac{\rho_{l}+\rho_{g}}{2\,\rho_c}\,,
	\end{equation} 
where $\rho_{l,g}$ are the densities of liquid and gas phases and $\rho_c$ is the critical density.  Different regions in the thermodynamic state space can be characterized with respect to the prevalence of hard-core repulsive or attractive contributions to pressure $P$ (we assume a common case of a simple fluid with central interparticle interaction potential $\Phi(r)$):
	\begin{equation}\label{eq:zp}
		Z = \frac{P}{\rho\,T} = 1-\frac{2\piup \,\rho}{3\,T}\int r^3 \,\frac{\partial\, \Phi(r)}{\partial\, r}\,g_2(r;\rho,T)\, \rd\,r\,,
	\end{equation}
where $T$ is the temperature of a fluid and $g_2$ is its correlation function. These regions can be labelled as soft ($Z<1$) and hard ($Z>1$) fluid regimes, correspondingly. The so-called Zeno-line:
	\begin{equation}\label{eq:z1}
		Z = \frac{P}{\rho\,T} = 1
	\end{equation}
serves as the separatrix between these regimes \cite{eos_zenobenamotz_isrchemphysj1990}. This line can be easily obtained for any equation of state (EoS) and tested via comparison with the available thermodynamic data \cite{eos_zenoapfelbaum_jpchemb2008}. For the van der Waals  (vdW) EoS, it is exactly straight, which was first noted by Batschinsky \cite{eos_zenobatschinski_annphys1906}:
\begin{equation} \label{eq:z1btschk}
Z = 1 \Rightarrow \frac{\rho}{\rho_{Z}}+\frac{T}{T_{Z}} = 1,
\end{equation}
with $T_Z=a/b\,,\,\,\rho_{Z}=1/b$ and $a,b$ are standard parameters of the vdW EoS \cite{book_ll5_en}. According to \cite{eos_zenobenamotz_isrchemphysj1990}:
	\begin{equation}\label{eq:tznz}
		B_{2}(T_{Z})= 0\,,\quad  \rho_Z= \frac{ T_Z }{B_3\left(\,T_Z\,\right)}\,\left. \frac{\rd B_2}{\rd T}\right|_{T= T_Z}\,,
	\end{equation}
where $B_{2,3}$ are corresponding virial coefficients. As thermodynamic data analysis shows \cite{eos_zenoapfelbaum_jpchemb2008}, the Zeno-line \eqref{eq:z1} has an approximate linear character for many real fluids.
	
Notably, beyond the fluctuational region where $\tilde{\rho}_d$ has a singular behaviour \cite{crit_diamerminrehr_prl1971,crit_singdiam_zhetp1972,crit_yydiamfisherorkoulas_prl2000}, the density diameter \eqref{eq:diam} also displays a linear temperature dependence known as the law of the rectilinear diameter (LRD) \cite{crit_diam1}:
	\begin{equation}\label{eq:lrd}
		\tilde{\rho}_{d} = 1+ A\,\left(1-\frac{T}{T_c}\right)\,\,, \quad A>0,
	\end{equation}
where $T_c$ is the critical temperature.
	
Both empirical regularities \eqref{eq:z1}, \eqref{eq:lrd} are very instrumental in locating critical points in numerical simulations \cite{eos_zenomietc_jcp2024} and in systems such as metals where they are beyond the reach of a real experiment \cite{liqmetals_zenoapfelbaum2_cpl2009,liqmetals_zenosapfvrbalkaline_jml2017,eos_zenolineionic_cpl2017}. It should be noted that while the LRD \eqref{eq:lrd} refers to the coexistence line, the Zeno-line regularity~\eqref{eq:z1} spans from supercritical fluid region to dense liquid region and lies well beyond the critical point ($Z_c \approx 0.3$). Thus, the concept which would encapsulate both these regularities must deal with the global nature of the fluid state. The ubiquity of these regularities in such different fluid systems, from simple fluids to liquid metals, hints at the underlying symmetry  which could serve as the basis for extended PCS. 

	The aim of this work is to present some recent results on the global isomorphism between simple fluid and the Ising-like model (lattice gas). This approach is based on a simple geometrical picture where the regularities \eqref{eq:diam} and \eqref{eq:z1btschk} appear as images of corresponding elements of the lattice gas phase diagram~\cite{eos_zenome0_jphyschemb2010,eos_zenomeunified_jphyschemb2011}. 
	
The paper is organized as follows. In section~\ref{sec:lgtri} we give a simple global isomorphism transformation picture using the triangle of liquid-gas states \cite{eos_zenotriapfelbaum_jpchemb2006} and its basic corollaries for the liquid-vapour equilibrium. This includes the relations between parameters of the transformation and critical point, the influence of quantum effects on the binodal asymmetry, and the relation between critical parameters of 3D and 2D fluids. In section~\ref{sec:microgi} we outline the microscopic nature of the global isomorphism between fluid and lattice Ising-like models. The results are discussed in the concluding section.
	
\section{Global isomorphism transformation and its parameters}\label{sec:lgtri}
\begin{figure}
			\centering
	\includegraphics[width=0.6\textwidth]{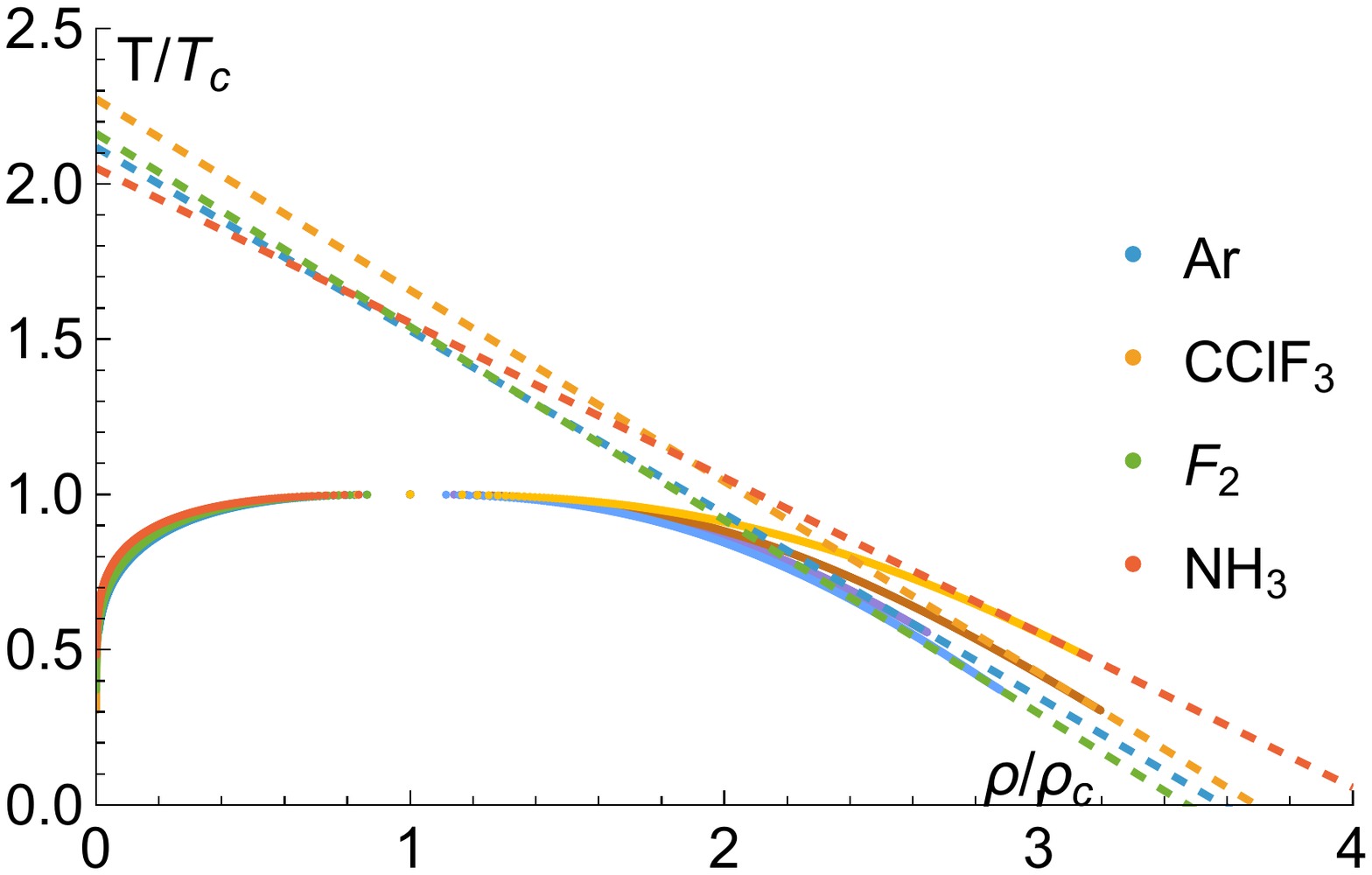}
		\caption{(Colour online) Tangent construction of the Zeno-element \eqref{eq:zenoelement} using NIST binodal data \cite{book_nist69} for some molecular fluids in table~\ref{tab:zgi}.
		}
		\label{fig:figtng}
	\end{figure}

The (approximate) linearity of the Zeno line and its parameters, augmented with its tangency to the liquid branch of the binodal at low temperature \cite{eos_zenoapfelbaum1_jpcb2009}, leads to the concept of a liquid-gas triangle of states into which the binodal is ``inscribed'' \cite{eos_zenotriapfelbaum_jpchemb2006}. In fact, in such a picture, one does not need the Zeno line itself with the parameters $T_Z$ and $\rho_Z$ defined by \eqref{eq:tznz} since its linearity is not a universal law. Instead, one can use what we call the Zeno element:
\begin{equation}\label{eq:zenoelement}
\frac{T}{T_*}+\frac{\rho}{\rho_{*}} = 1\,,
\end{equation}
with $T_*$ and $\rho_{*}$ as the corresponding parameters of this line due to its construction (see figure~\ref{fig:figtng}). Thus, the line \eqref{eq:zenoelement} is defined directly as the tangent to the analytical continuation of the liquid branch of the binodal to the low-temperature region $T\to 0$. Such a continuation is possible via mapping the symmetric binodal of the Ising-like model of lattice gas with the Hamiltonian
\begin{equation}\label{eq:hamlg}
H = -\sum\limits_{\left\langle\, ij \,\right\rangle}\,J_{ij} \, n_{i}\,n_{j} - h \,\sum\limits_{i}\,n_{i}
\end{equation}
onto the symmetrized liquid-gas binodal of the fluid \cite{eos_zenome0_jphyschemb2010}. In \eqref{eq:hamlg}, $J$ is the interaction energy, $n_{i} = 0,1$ is the site occupation variable and $h$ is the chemical potential. Assuming LRD \eqref{eq:lrd}, the mapping takes a simple form: 
\begin{equation}\label{eq:projtransfr}
\rho =\, \rho_{*}\,\frac{x}{1+z \,\tilde{t}}\,,\quad
T =\, T_*\,\frac{z\, \tilde{t}}{1+z \,\tilde{t}}\,,
\end{equation}
of the projective transformation with $z$ being the skewness parameter. Here, $x =\langle n_i \rangle$ is the lattice gas density and $\tilde{t} = t/t_c$ is its temperature $t$ normalized to the critical one $t_c$.
Such a transformation defines the correspondence between two pairs of concurrent triples of straight lines:
\begin{eqnarray*}
\text{ideal gas line:}&\,\,\rho = 0\,\,&\Rightarrow\, x= 0\,,\\
\text{binodal diameter:}&\,\,\tilde{\rho}_d\,\,&\Rightarrow\, x= 1/2\,,\\
\text{Zeno-element:}&\,\,\frac{T}{T_*}+\frac{\rho}{\rho_{*}} = 1\,\,&\Rightarrow \, x= 1\,.
\end{eqnarray*}
Therefore, from \eqref{eq:projtransfr}, the critical temperature and density are as follows:
	\begin{equation}\label{eq:nctc}
		\rho_c=\frac{\rho_{*}/2}{1+z}\,,\quad T_c = T_{*}\frac{z}{1+z}.
	\end{equation}
Density parameter $\rho_*$ represents the highest density state associated with the characteristic temperature $T_{*}$, and, for model systems with a given interaction potential, it can be estimated as:
	\begin{equation}\label{eq:rhostar}
		\rho_{*}= \frac{ T_*}{B_3\left(\,T_*\,\right)}\,\left. \frac{\rd\,B_2}{\rd T}\right|_{T= T_*}\,.
	\end{equation}
	\begin{table}[!h]
		\centering
		\renewcommand{\arraystretch}{0.1}
		\caption{Checking the relations \eqref{eq:zzz} and \eqref{eq:2nctc} for molecular fluids using NIST data \cite{book_nist69}.}\label{tab:zgi}
		\vspace{0.3cm}
		\begin{tabular}{|c|c|c|c|c|c|c|}
			\hline  
		\strut	\text{Fluid} & $z_T=\frac{T_c}{T_*-T_c}$ & $z_{\rho }=\frac{\rho _*}{2\rho _c}-1$ & $z=\frac{\rho _*}{2\rho _c}\frac{T_c}{T_*}$ & $2\frac{\rho _c}{\rho_*}+\frac{T_c}{T_*}=1$ & $T_*/T_c$ & $\rho_*/\rho_c$ \\
			\hline\strut
			$\text{CO}_2$ & 1.23 & 1.08 & 1.15 & 1.03 & 1.81 & 4.15 \\
			\hline\strut
			C$_{12}$H$_{26}$& 1.04 & 1.14 & 1.09 & 0.98 & 1.96 & 4.28 \\
			\hline\strut
			C$_{10}$H$_{22}$& 1.03 & 1.06 & 1.04 & 0.99 & 1.97 & 4.12 \\
			$\text{SF}_6$ & 1.12 & 0.97 & 1.04 & 1.04 & 1.9 & 3.94 \\
			\hline\strut
			$\text{NH}_3$ & 0.95 & 1.06 & 1. & 0.97 & 2.05 & 4.11 \\
			\hline\strut
			$\text{PhCH}_3$ & 1.03 & 0.97 & 1. & 1.02 & 1.97 & 3.94 \\
			\hline\strut
			N$_2$O & 1.03 & 0.95 & 0.99 & 1.02 & 1.97 & 3.9 \\
			\hline\strut
			C$_6$H$_{14}$ & 0.99 & 0.98 & 0.98 & 1. & 2.01 & 3.96 \\
			\hline\strut
			$\text{SO}_2$ & 0.96 & 0.99 & 0.98 & 0.99 & 2.04 & 3.98 \\
			\hline\strut
			C$_6$H$_{12}$ & 0.99 & 0.93 & 0.96 & 1.01 & 2.01 & 3.87 \\
			\hline\strut
			C$_5$H$_{12}$& 0.95 & 0.93 & 0.94 & 1.01 & 2.05 & 3.86 \\
			\hline\strut
			C$_6$H$_6$ & 0.91 & 0.89 & 0.9 & 1. & 2.1 & 3.79 \\
			\hline\strut
			$\text{Ne}$& 1. & 0.79 & 0.9 & 1.06 & 2. & 3.59 \\
			\hline\strut
			$\text{CCl}_2\text{F}_2$ & 0.89 & 0.89 & 0.89 & 1. & 2.13 & 3.77 \\
			\hline\strut
			$\text{CO}$ & 0.92 & 0.85 & 0.89 & 1.02 & 2.09 & 3.71 \\
			\hline\strut
			H$_2$S & 0.9 & 0.87 & 0.89 & 1.01 & 2.12 & 3.75 \\
			\hline\strut
			$\text{Kr}$ & 0.9 & 0.82 & 0.87 & 1.02 & 2.11 & 3.65 \\
			\hline\strut
			$\text{Xe}$ & 0.9 & 0.83 & 0.86 & 1.02 & 2.12 & 3.65 \\
			\hline\strut
			C$_4$H$_{10}\text{-iso}$ & 0.84 & 0.88 & 0.86 & 0.99 & 2.19 & 3.76 \\
			\hline\strut
			Ar& 0.91 & 0.8 & 0.86 & 1.03 & 2.1 & 3.6 \\
			\hline\strut
			C$_4$H$_{10}$ & 0.83 & 0.88 & 0.85 & 0.98 & 2.21 & 3.76 \\
			\hline\strut
			CFCl$_3$ & 0.8 & 0.89 & 0.84 & 0.98 & 2.24 & 3.77 \\
			\hline\strut
			$\text{COS}$ & 0.86 & 0.82 & 0.84 & 1.01 & 2.17 & 3.63 \\
			\hline\strut
			NF$_3$ & 0.78 & 0.9 & 0.83 & 0.97 & 2.28 & 3.8 \\
			\hline\strut
			CClF$_3$ & 0.8 & 0.85 & 0.82 & 0.98 & 2.25 & 3.7 \\
			\hline\strut
			N$_2$ & 0.84 & 0.8 & 0.82 & 1.01 & 2.18 & 3.59 \\
			\hline\strut
			F$_2$ & 0.86 & 0.74 & 0.8 & 1.04 & 2.16 & 3.47 \\
			\hline\strut
			CH$_4$ & 0.81 & 0.76 & 0.79 & 1.02 & 2.23 & 3.53 \\
			\hline\strut
			C$_2$H$_6$ & 0.73 & 0.8 & 0.76 & 0.98 & 2.36 & 3.6 \\
			\hline\strut
			CF$_4$ & 0.72 & 0.78 & 0.75 & 0.98 & 2.39 & 3.57 \\
			O$_2$ & 0.7 & 0.75 & 0.72 & 0.98 & 2.42 & 3.5 \\
			\hline\strut
			para-H$_2$ & 0.46 & 0.41 & 0.44 & 1.02 & 3.19 & 2.83 \\
			\hline\strut
			H$_2$ & 0.44 & 0.41 & 0.43 & 1.01 & 3.26 & 2.83 \\
			\hline\strut
			He & 0.09 & 0.09 & 0.09 & 1. & 12.23 & 2.17 \\
			\hline
		\end{tabular}  
	\end{table}

\begin{figure}[!t]
		\centering
		\begin{subfigure}[b]{0.45\textwidth}
			\centering
			\includegraphics[width=\textwidth]{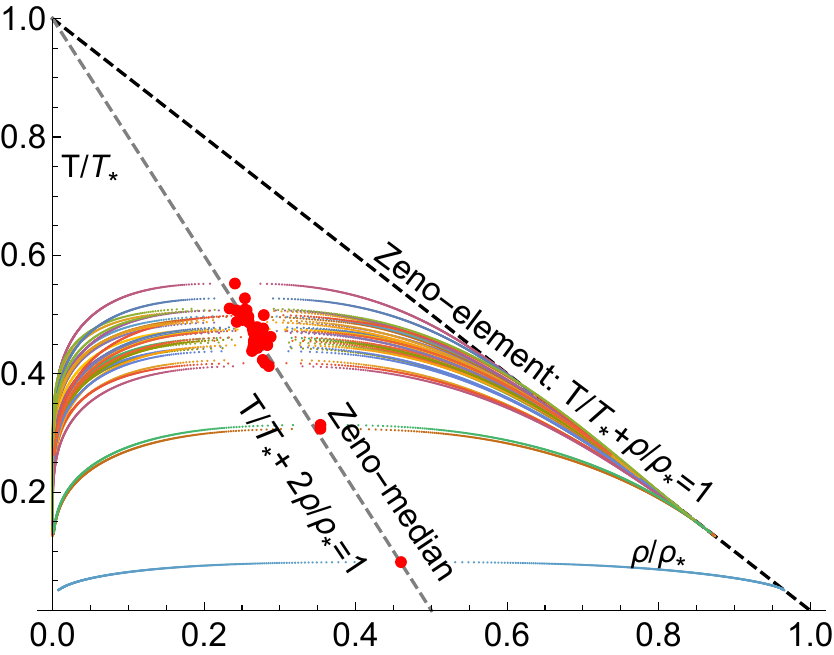}
		\end{subfigure}
		\hspace{0.5cm}
		\begin{subfigure}[b]{0.45\textwidth}
			\centering
		\includegraphics[width=\textwidth]{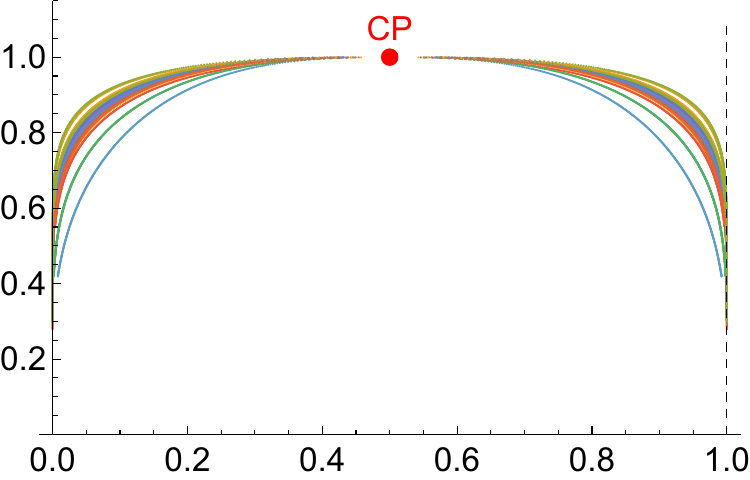}
		\end{subfigure}
		\caption{(Colour online) NIST binodal data \cite{book_nist69} for molecular fluids listed in table~\ref{tab:zgi} (a) reduced to $T_{*},\rho_{*}$ and (b) symmetrized with respect to diameter. Critical points (CP) are marked with red.}
		\label{fig:nistbin}
	\end{figure}

	In view of global isomorphism, $\rho_{*}$ corresponds to a fully occupied state $x=1$ of the lattice gas. This defines the number of cells $\mathcal{N} = \rho_{*}\,V$ for the isomorphic  lattice gas model in a volume $V$ of a fluid. Further, we use conventional dimensionless units for the temperature $T\to T/\varepsilon$ and the density $\rho\to \rho\sigma^{3}$, where $\varepsilon$ is the characteristic energy scale of the interaction, e.g., the minimum of the potential, and $\sigma$ is the hard-core diameter of a particle.
	In such a picture for real fluids, the Zeno-element \eqref{eq:zenoelement} can be  easily constructed (see figure~\ref{fig:figtng}). The fluid binodal rescaled with the Zeno-element parameters $T_*,\rho_*$  instead of common critical parameters $T_{c},\rho_c$ explicitly displays regularities (see figure~\ref{fig:nistbin}). In the global isomorphism approach, $T_*$ can be determined as the Boyle temperature in the vdW approximation:
	\begin{equation}\label{eq:tstar}
		B^{(\rm vdW)}_2(T_{*}) = 0\,,\quad  T_{*} =  T^{(\rm vdW)}_{B}  = \frac{a}{b}\,,
	\end{equation}
	with
	\begin{equation} \label{eq:vdw_ab}
		a =\,\, -2\piup\,\int\limits_{\sigma}^{+\infty}\Phi_{\text{attr}}(r)\,r^2\,\rd r\,.    
	\end{equation}
	Here, $\Phi_{\text{attr}}(r)$ is the attractive part of the interaction potential $\Phi(r)$, $\sigma$ is particle diameter so that $b = \frac{2\piup}{3}\,\sigma^{3}$. The reason for using \eqref{eq:rhostar} and \eqref{eq:tstar} instead of standard virial ones \eqref{eq:tznz} as the parameters of the transformation \eqref{eq:projtransfr} comes from the very idea of the fluid-lattice gas global isomorphism. By the definition, $T_{*}$ includes the attractive part of the potential only in coherence with the structure of the lattice gas  Hamiltonian \eqref{eq:hamlg}. 
	
	Obviously, the following relations:
	\begin{equation}\label{eq:zzz}
		z = \frac{\rho_{*}\,T_c}{2\,\rho_{c}\,T_*}= \frac{T_{c}}{T_{*} - T_{c}}= \frac{\rho_{*}}{2\,\rho_{c}} -1,
	\end{equation}
	are direct inferences of \eqref{eq:projtransfr} and therefore the following relation is fulfilled:
	\begin{equation}\label{eq:2nctc}
		2\,\frac{\rho_{c}}{\rho_{*}}+\frac{T_c}{T_*}=1.
	\end{equation}
	Note that \eqref{eq:2nctc} does not depend on $z$ and this directly can be tested using the available thermodynamic data to confirm the validity of \eqref{eq:projtransfr} and tangent construction for finding density parameter $\rho_{*}$. The results  are presented in table~\ref{tab:zgi} and demonstrate that \eqref{eq:2nctc} takes place with good accuracy. 
	
\begin{figure}
		\centering
		\begin{subfigure}[b]{0.47\textwidth}
			\centering
			\includegraphics[width=\textwidth]{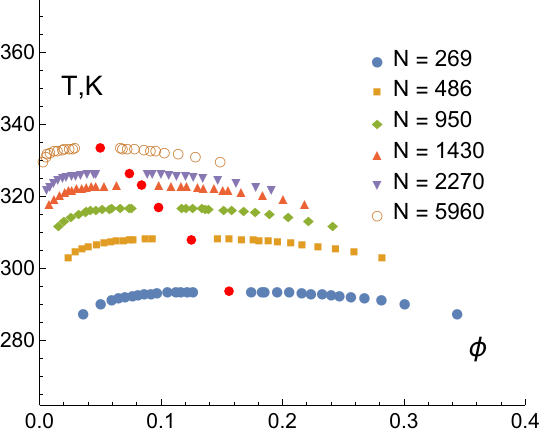}
			\caption{}
		\end{subfigure}
		\hspace{0.15cm}
		\begin{subfigure}[b]{0.47\textwidth}
			\centering
			\includegraphics[width=\textwidth]{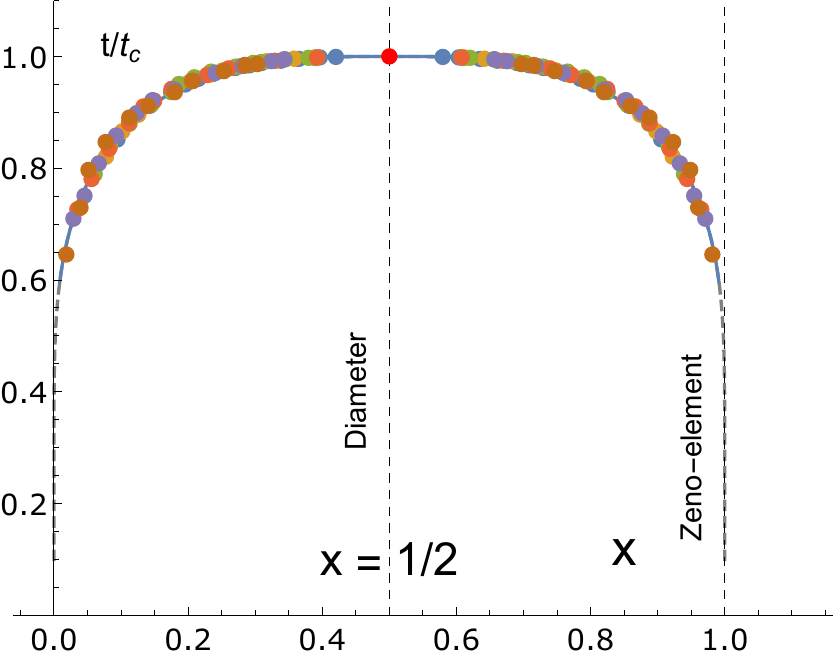}
			\caption{}
			\label{fig:floryxiabinz}
		\end{subfigure}\\
		\begin{subfigure}[b]{0.47\textwidth}
			\centering
			\includegraphics[width=\textwidth]{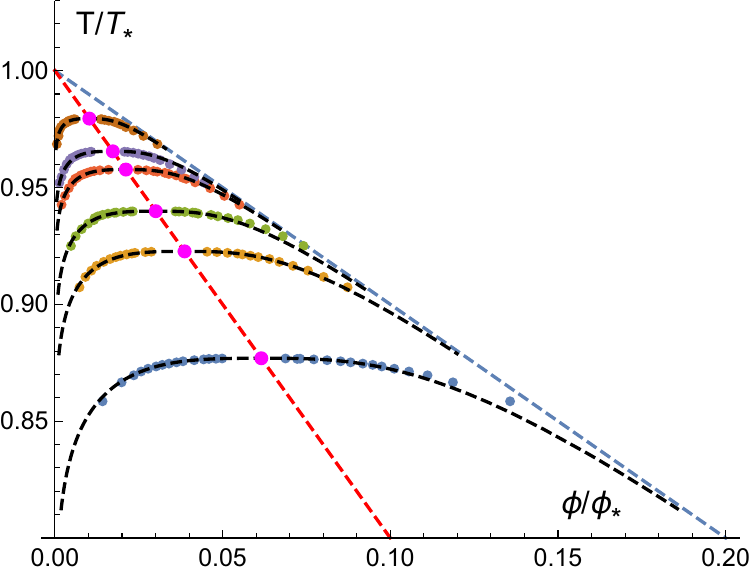}
			\caption{}
			\label{fig:xiadatsymm}
		\end{subfigure}
		\hspace{0.1cm}
		\begin{subfigure}[b]{0.47\textwidth}
			\includegraphics[width=\textwidth]{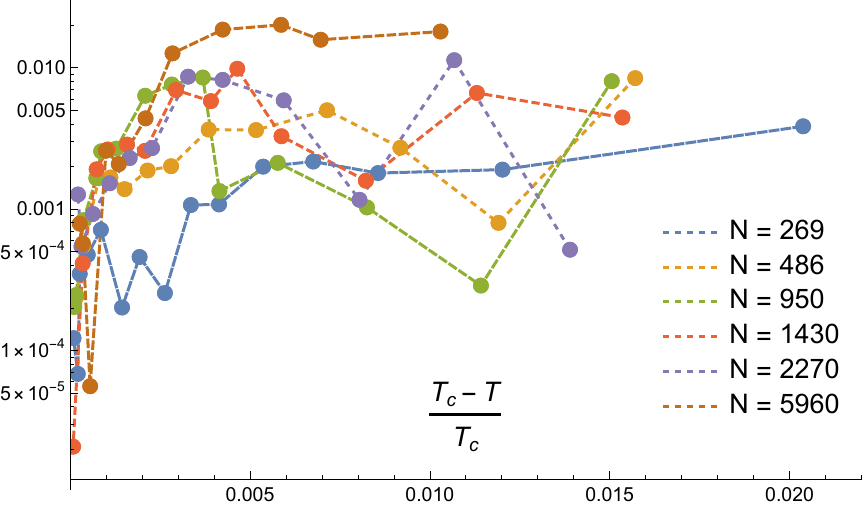}
			\caption{}
		\end{subfigure}
		\caption{(Colour online) (a) Coexistence curves of \cite{crit_florypolymersymmxia_jcp1996} (b) mapped onto 3D Ising model binodal (c) coexistence curves (a) scaled in coordinates $T/T_*, \phi/\phi_*$ and (d) absolute deviations of the mapped 3D Ising binodal from the experimental data.}
		\label{fig:floryzz}
	\end{figure}

	Additional argument in favour of \eqref{eq:tstar} comes from the comparison of $T_*$ in \eqref{eq:nctc} with the Flory $\theta$-point in a dilute polymer solution where the binodal itself degenerates into triangle in $T-\phi$ plane with $\phi$ being the polymer volume fraction \cite{book_floryppchem_1953}. It is clear that $T_c\to T_*$ as $z\to \infty$. Therefore, an analogy appears between \eqref{eq:nctc} and the Flory-Huggins theory result for the critical temperature and $\theta$-point of polymer solutions  \cite{book_floryppchem_1953}:
	\begin{equation}\label{eq:florytemp}
		T_c = \frac{\theta}{\left(1+\frac{1}{\sqrt{N}}\right)^2} \,,\quad \phi_c = \frac{1}{1+\sqrt{N}},
	\end{equation}
	where $N$ is the degree of polymerization.
	Indeed, comparison  \eqref{eq:nctc} and \eqref{eq:florytemp} within such an approximation leads to a correspondence: 
	\begin{equation}\label{eq:florytz}
		T_{*}/\theta =  1 + O\left(N^{-1/2}\right)\,,\quad z \propto\sqrt{N}/2\,,
	\end{equation}
	and obviously we have:
	\begin{equation}\label{eq:thetstar}
		\theta_{*} = \lim\limits_{N\to \infty} T_{*}.
	\end{equation}
We consider this result as a pretty natural one within our approach, since $\theta$-temperature is defined in a way similar to $T_{*}$ (see, e.g., \cite{book_polymers_oxp1990}). Here, we use \eqref{eq:projtransfr} to directly demonstrate how the data of \cite{crit_florypolymersymmxia_jcp1996} for binodals of polymer solutions can be represented within the concept of the triangle of liquid-gas states used for molecular liquids above. At first, we symmetrize the binodals with respect to diameters (see figure~\ref{fig:xiadatsymm}) and map them onto the 3D Ising model (simple cubic lattice data of \cite{crit_3disingliufisher_physa1989} were used) binodal, thus finding $z_{N}$ parameter of the projective transformation for each degree of polymerization $N$. Note that this symmetrization is exact and does not use any fitting parameters. Finally, the symmetrized binodals are mapped back into the original coordinates (see figure~\ref{fig:floryxiabinz}) scaled with the corresponding values $T_*$ and $\rho_{*}\leftrightarrow \phi_{*}$ as the analogue of $\rho_{*}$:
	\begin{equation}\label{eq:tnznphin}
		T_* = T_c(N)\,\frac{1+z_{N}}{z_{N}}\,,\quad  \phi_{*} = 2\,\phi_c\,(1+z_{N}),
	\end{equation} 
for each $N$. The results of data fitting are in table~\ref{tab:zpoly}. Such a procedure differs from  nonlinear redefinition of the order parameter:
	\begin{equation}\label{eq:sancezpsi}
		\psi  = \frac{\phi}{\phi+R_c\,(1-\phi)},
	\end{equation}
used in \cite{crit_florypolymersymmsanchez_jap1985}, where $R_c$ is the adjusted parameter of symmetrization. 
Moreover, in our approach, the data fit the universal curve (see figures~\ref{fig:floryzz}\,c, d) with clear physical meaning in lieu of a pure empirical fitting curve in terms of parameter $\psi$ \eqref{eq:sancezpsi} \cite{crit_florypolymersymmxia_jcp1996}. Flory $\theta$-temperature was determined using relation \eqref{eq:2nctc}.
	\begin{table}[htb]
		\caption{Values of $z_N$ according to data fitting with the use of \eqref{eq:tnznphin}.}
		\label{tab:zpoly}
		\begin{center}
			\renewcommand{\arraystretch}{1.2}
			\begin{tblr}{vlines, cells = {c,m},
					cell{4}{2} = {c=6}{}, 
				}
				\hline
				${N}$ & 269 & 486 & 950 & 1430 & 2270 & 5960 \\
				\hline
				${z_{N}}$ & 7.12 & 11.93 & 15.62 & 22.62 & 27.95 & 47.71 \\
				\hline
				${\phi _*}$ & 2.53 & 3.23 & 3.26 & 3.97 & 4.28 & 4.87 \\
				\hline
				${\theta}, \, ^{\circ}$C& 66.1&  &  &  &  & \\
				\hline
			\end{tblr}
		\end{center}
	\end{table}

\subsection{Quantum corrections}\label{subsec:qcorr}
From a physical point of view, the quantum delocalization due to particle-wave duality being taken into account should lead to a more symmetric liquid-gas binodal because this effect tends to reduce the  difference between ``particles'' and ``holes''. Experimental data for real fluids demonstrate this clearly (see table~\ref{tab:slopefluid}) with helium as a canonical example of a quantum fluid for which the diameter slope is almost an order of magnitude less than the one for other fluids. 
	\begin{table}[!t]
		\centering
		\caption{Diameter slope for different fluids (data from \href{https://webbook.nist.gov/chemistry/fluid/}{NIST database}).}
		\vspace{0.3cm}
		\label{tab:slopefluid}
		\begin{tabular}{|c|c|c|c|c|c|c|c|c|}
			\hline
			Fluid & He &para-H$_2$& H$_2$ & D$_2$ & CH$_3$OH &   Ne & F$_2$ & Ar \\
			\hline
			Slope & 0.10 & 0.39 & 0.40 & 0.50 & 0.57 &  0.67 & 0.70 & 0.74 \\
			\hline
			\hline
			Fluid: & CH$_4$ &O$_2$& N$_2$ & Kr & Xe &   CO & CO$_2$ & H$_2$O \\
			\hline
			Slope: & 0.74 & 0.76 & 0.77 & 0.78 & 0.79 &  0.82 & 0.95 & 0.96 \\
			\hline
		\end{tabular}
	\end{table}
In this section we  quantitatively demonstrate in the first order on the delocalization parameter:
$$\varkappa =  (\lambda(T_*)/\sigma)^3 \ll 1\,,\quad \lambda(T) = \frac{h}{\sqrt{2\piup\,m\,T}}\,.
$$
Our main result is that such a delocalization  increases the ratio $S_*=T_*/\rho_*$ which is the skewness of the liquid-gas triangle. This makes the binodal more symmetric. To demonstrate this, we assume Bose-Einstein statistics, which is obviously the case here. Thus, expanding $T_*, \rho_*$ in quantum parameter $\varkappa$, we can write:
	\begin{equation}
		\label{eq:tz_nz_quant}
		T_*/T^{(0)}_* = 1+T_q\,\varkappa +o(\varkappa)\,,\quad    \rho_*/\rho^{(0)}_* 
		= 1+\rho_q\,\varkappa +o(\varkappa)\,\,,
	\end{equation}
with $T_{q}$ and $\rho_{q}$ as expansion coefficients on the quantum delocalization parameter $\varkappa$ (subscript ``q'' stands for ``quantum''). Here, $T^{(0)}_*$ is the classical Zeno temperature and $\rho^{(0)}_*$ is the classical Zeno density, which are defined by the equations \eqref{eq:tstar} and \eqref{eq:rhostar}, respectively. 
The change of $T_*$ is due to  quantum correlations for an ideal Bose gas which lead to an effective attraction \cite{book_balesku_statmech}. The change of the density parameter $\rho_{*}$ is determined by the corresponding shift of $T_*$ in \eqref{eq:rhostar}. Thus, 
	\begin{align}
		S_*/S^{(0)}_* =& \, 1+s_q \,\varkappa +o(\varkappa),\nonumber\\
		s_q=&\, T_q-\rho_q.
			\label{eq:tz_nz_quant1}
	\end{align}
In case of the Lennard-Jones potential, 
	\begin{equation}
		\label{eq:lj}
		\Phi_{\text{LJ}}(r)=4\,\varepsilon\, \Big(\Big(\frac{\sigma}{r} \Big)^{12} -\Big(\frac{\sigma}{r} \Big)^6 \Big)\,,
	\end{equation}
the calculation gives:
	\begin{equation}
		\label{eq:sqlj}
		S_*/S^{(0)}_* = 1+1.03\,\varkappa\,\,.
	\end{equation}
This indicates the tendency to symmetrization of the binodal caused by quantum delocalization.
\subsection{Application to 2D fluid case}\label{subsec:2d}
Two-dimensional fluid systems are crucial for testing the global isomorphism (GI) predictions in view of numerous exact results for lattice systems \cite{book_baxterexact}. Surprisingly, the binodal of liquid-gas equilibrium in a 2D fluid system with the corresponding critical exponent of the Ising model is hard to obtain analytically within the general integral equation approach of liquid state theory \cite{crit_2dljpy_jcp977,eos_scoza2dlattgas_jcp1998}.    
	
The Onsager solution of 2D Ising model \cite{crit_onsager_pr1944} with the binodal equation:
	\begin{equation}\label{eq:isingbinodal2d}
		t(x)/t_c = \frac{{\rm arcsinh}(1)}{{\rm arcsinh}\left((1 - (2 x - 1)^{8})^{-1/4}\right)}\,,\quad t_c = \frac{2}{{\rm arcsinh}(1)}
	\end{equation}
can be used to obtain information for planar fluid systems, both the model and the real ones. The latter can be obtained as the adsorbed molecular monolayers of different fluids on graphite substrate \cite{eos_2dljabraham_physrep1981,eos_2dargongraphit_collsurf2021}. Now, we demonstrate how one can establish the relation between critical parameters of 3D fluid and its 2D monolayer  using global isomorphism relations on the basis of the previous results \cite{crit_globalisome_jcp2010}. We use MC simulation data on Ar and Xe films on graphite substrate \cite{eos_2darxe_condmat2012} where the Lennard-Jones potential \eqref{eq:lj}  was used with the following parameters: 
$$u^{(\text{Ar})}_0 = 120\,\text{K}\,,\quad\sigma_{\text{Ar}}  = 3.4\,\text{\AA}\,, \quad u^{(\text{Xe})}_0 = 221\, \text{K}\,, \quad\sigma_{\text{Xe}}  = 4.1\,\text{\AA}\,.$$
For the LJ potential in 2D according to \cite{crit_globalisome_jcp2010}, the theoretical values for parameters of \eqref{eq:projtransfr} and the critical point are:
	\begin{equation}\label{eq:tz}
		z_{\text{2D}}=1/3\,,\quad T_{*}/\varepsilon = 2\,,\quad \rho_{*}\approx 0.94\,, \quad  T_{c} = 0.5\,,\quad \rho_{c} \approx 0.353\,.
	\end{equation}
We substitute \eqref{eq:isingbinodal2d} into \eqref{eq:projtransfr} with $T_*$ fixed but treating $z$ and $\rho_{*}$ as fitting parameters. 
	\begin{figure}[!t]
		\centering
		\begin{subfigure}[b]{0.45\textwidth}
			\centering
			\includegraphics[width=\textwidth]{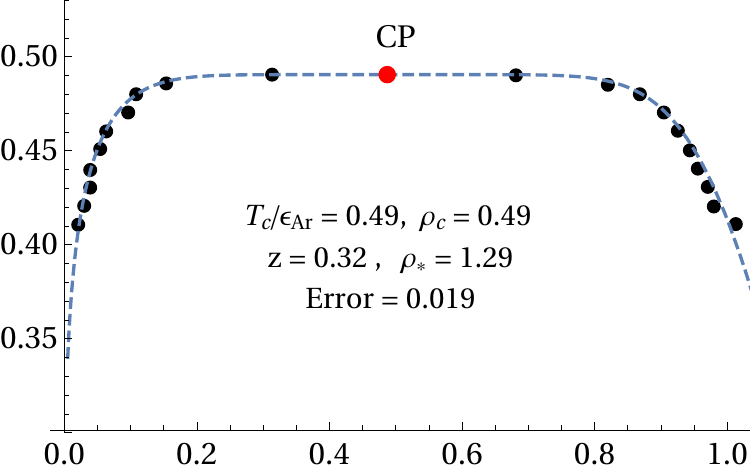}
			\caption{}
			\label{fig:figbintri}
		\end{subfigure}
		\hspace{0.5cm}
		\begin{subfigure}[b]{0.45\textwidth}
			\centering
		\includegraphics[width=\textwidth]{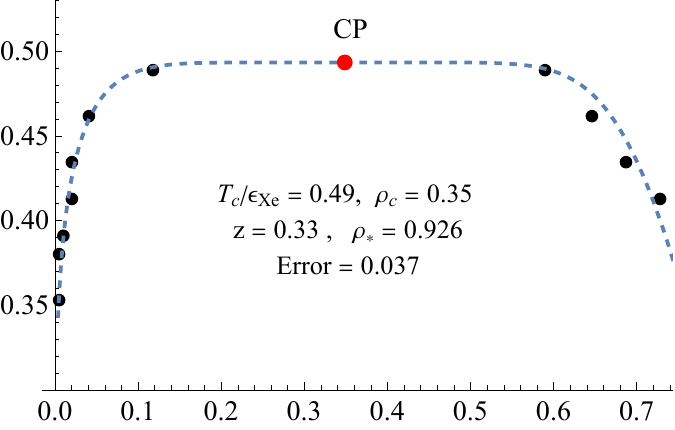}
			\caption{}
			\label{fig:figbinz}
		\end{subfigure}
		\caption{(Colour online) Binodal data for  (a) argon and (b) xenon processed by mapping \eqref{eq:isingbinodal2d} with~\eqref{eq:projtransfr} (dashed line).}
		\label{fig:2dmonoar}
	\end{figure}
Thus, the parametric representation for the binodal of LJ fluid is obtained and compared with the known numerical simulation results (see figure~\ref{fig:2dmonoar}). Now, it is easy to relate the bulk-monolayer fluid critical temperature for Lennard-Jones fluid with the value for the bulk phase: $T_{2c} = 3/8\,T_{3c}$. 
	\begin{table}[!t]
		\caption{Critical temperatures for monolayers modelled as 2D LJ fluid.}
		\label{tab:z2dmono}
		\begin{center}
			\renewcommand{\arraystretch}{1.2}
			\begin{tblr}{vlines, cells = {c,m},
				}
				\hline
				&Ar& Kr & Xe\\
				\hline
				 3D: $T_c^{(\text{exp})}$, K\,, \cite{book_nist69} & $150.8$& $209.48$ & $289.7$\\
				\hline
				 2D: $T_c^{(\text{exp})}$, K\,, \cite{eos_2dnoble_physrev1979} & $59$& $86$& $117$\\
				\hline
				\hline
				 2D: $T_c^{(\text{GI})}$, K & $57$& $79$& $109$\\
				\hline
			\end{tblr}
		\end{center}
	\end{table}
This gives a rather good agreement  with the simulation data (see table~\ref{tab:z2dmono}) though the values are lower by $\sim 5\div 8\%$ \cite{eos_2dnoble_physrev1979}. This can be attributed to the difference between real and model interaction potentials \cite{eos_kryptonlj_jcplett2022}. 

Another important property of liquid-gas equilibrium, for which our approach may be applied, is the surface tension. Based on the relation between thermodynamic potentials of the LJ fluid and the lattice gas \cite{eos_zenomeunified_jphyschemb2011}, the following correspondence between the surface tensions of the LJ fluid and the lattice model takes place:
	\begin{equation}\label{eq:st_iso}
		\gamma_{\text{LJ}}(T)= \gamma_{\text{latt}}(t(T))\,,\quad 
		t(T) = \frac{t_c}{z}\,\frac{T}{T_*-T} ,
	\end{equation}
provided that geometrical sizes of the corresponding systems are the same. 

In the following analysis we use the dimensionless units for the surface tension: $s_{\text{latt}}=\gamma_{\text{latt}} \,l/J$, where $J$ is the interaction constant of the Ising model and $l$ is the lattice spacing. The surface tension of the LJ fluid is measured in the corresponding units: $s_{\text{LJ}}=\gamma_{\text{LJ}} \sigma/\varepsilon$, where $\varepsilon$ and $\sigma$ are the parameters of the LJ potential~\eqref{eq:lj}.
	
\begin{figure}[!t]
		\centering
	\includegraphics[width=0.7\linewidth]{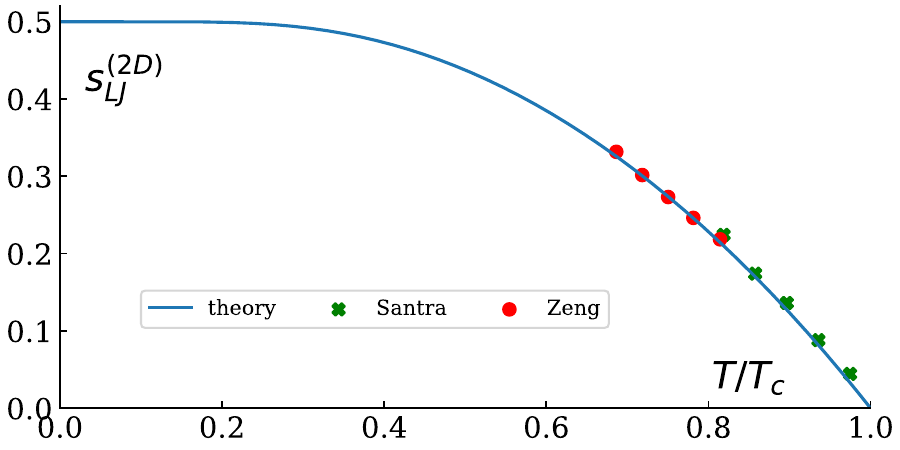}
		\caption{(Colour online) Data of \cite{crit_2dsurftens_jcp1996,liq_2dsurftens_jcp2009} and theoretical curve \eqref{eq:st_2disingfit}.}
		\label{fig:figsurf2dljall}
	\end{figure}

In the context of the global isomorphism, equation \eqref{eq:st_iso} allows one to get the information about the surface tension of the LJ fluid on the basis of the corresponding lattice model. In particular, for the surface tension of the 2D LJ fluid using the Onsager's solution \cite{crit_onsager_pr1944}, we have:
	\begin{equation}\label{eq:st_2disingfit}
		s^{(\text{2D})}_{\text{LJ}}(T) = \frac{1}{4}\,\left(2 + t(T)\,\ln
		\left(\,\tanh\,\frac{1}{t(T)} \,\right)\right)\,.
	\end{equation}
The factor $1/4$  appears due to the redefinition of the interaction constant in the lattice gas with respect to the Ising model \cite{book_baxterexact}. From \eqref{eq:st_2disingfit} and \eqref{eq:tz}, we get the following critical asymptotic for the surface tension:
	\begin{equation}\label{eq:critamplit_st}
		s^{(\text{2D})}_{\text{LJ}} = s^{(0)}_{\text{LJ}}\,|\tau|+o(\tau)\,,\quad \tau = 1-T/T_c,
	\end{equation}
and its critical amplitude is $s^{(0)}_{\text{LJ}} = 4/3$. We check the validity of our approach by applying equation~\eqref{eq:st_2disingfit} to the results of molecular simulations \cite{liq_2dsurftens_jcp2009} using $T_{*}$ and $z$ as the fitting parameters. The result is shown in figure~\ref{fig:figsurf2dljall}. We see that the simulation data agree quite well with the predictions of the global isomorphism approach. In addition, the value of the critical amplitude in the 2D case, $s^{(0)}_{\text{LJ}} \approx 1.318$ which follows from the simulations \cite{liq_2dsurftens_jcp2009}, is also in good agreement with the theoretical value, $s^{(0)}_{\text{LJ}} = 4/3$. We can conclude that the application of the global isomorphism to the surface tension of the 2D LJ fluid gives much better results than the other theoretical approaches used to analyze the results of the simulations \cite{liq_2dsurftens_jcp2009}, since they fail to reproduce both the binodal and the surface tension data. 

\section{Microscopic nature of global isomorphism}\label{sec:microgi}
Here, we outline the possibility of reducing the consideration of continuum fluid to the ``lattice'' model, leaving the details for further studies.
	
Let us consider the starting point of statistical theory --- the partition function:
	\begin{equation}\label{eq:bigpartfun}
		\Xi_{\text{Fluid}} = \sum_{N}\,\frac{\re^{\beta\,\mu\,N}}{N!}\int \re^{-\beta\,\sum\limits_{i<j}^{N}\Phi(r_{ij})}\,\prod_{i,j} \rd\mathbf{r}_{ij}\,.
	\end{equation}
The coordinate variables here are the dummy ones and particle configuration is completely encoded into the values $\Phi_{ij}$ of the potential. Therefore, we can consider the fixed lattice structure with the distribution of interaction constants induced by the disorder in particle coordinates. This way one can represent the grand partition function of a fluid in terms of the lattice model with additional averaging on the ``site-site'' interaction constants. Indeed, \eqref{eq:bigpartfun} can be written as:
	\begin{equation}\label{eq:bigpartfun0}
		\Xi(T,\mu,V) = \Xi_{0}(T,\mu,V) \, \Xi_{\text{int}}\,,
	\end{equation}
	where 
	$$
	\Xi_{0}(T,\mu,V) =	\sum\limits_{N}\,V^{N}\frac{\re^{\beta \mu  N}}{N!}\,,\quad 
	\Xi_{\text{int}} = \int \re^{\beta\sum\limits_{\left\langle\, i,j \,\right\rangle}\,\mathcal{J}_{ij}}\,\,p_{N}\left(\,\{\mathcal{J}\}\right)\,
	\rd\mathcal{J}\,.
	$$
The probability distribution for interaction values is:
	\begin{equation}\label{eq:pjdist}
		p_{N}\left(\,\{\mathcal{J}\}\,\right) = \int \prod\limits_{i,j}\,\delta\left(\,\Phi(r_{ij}) + \mathcal{J}_{ij}
		\,\right)\,\rd\Gamma_N\,,\quad \rd\Gamma_{N} = \frac{1}{V^N}\prod\limits^{N}_{i}\,\rd\mathbf{r}_i.
	\end{equation}
Note that this does not depend on the temperature though the dependence on the volume is present. It seems natural that in view of thermodynamic limit $N\gg1 , N/V = n$ it is reduced to the dependence on the particle density only. Certainly, the exact determination of distribution  $p_n$ in general is not possible, although in one dimensional case this should not be a problem. Now, we can go a little bit further and split the distribution of absolute values of $\mathcal{J}_{ij}$ from their signs by introducing the discrete variables: 
	\begin{equation}\label{eq:jtau}
		\mathcal{J}_{ij} = J_{ij}\,\tau_{i,j}\,,\quad \tau_{i,j} = {\rm sign}\,\mathcal{J}_{i,j} =  \pm 1 \,, \quad
		\left\vert\sum\limits_{j}\,\tau_{i,j} \right\vert\leqslant N-1 \notag.
	\end{equation}
This very idea appeared to be very useful in the theory of bind-random Ising model and spin glasses, though the analytical situation therein is easier because bond interactions $J_{ij}$ are considered as independent variables \cite{crit_nishimori_ptp1981,crit_nishimori_ptp1986}. Now, we may integrate over $J$-configuration: 
	\begin{equation}\label{eq:gtau}
		\mathcal{G}_{N}[\{\tau\};\beta,n] =	\int
		\re^{\beta\sum\limits_{\left\langle\, i,j \,\right\rangle}\,J_{ij}\,\tau_{i,j}
		}\,\,p_{N}\left(\,\{\mathcal{J}\}\right)\,
		\rd\,J   = \left\langle \re^{\beta\sum\limits_{\left\langle\, i,j \,\right\rangle}\,J_{ij}\,\tau_{i,j}}\right\rangle_{J} ,
	\end{equation}
considering \eqref{eq:gtau} as the characteristic functional of $\tau$-distribution, so the  $\tau$-Hamiltonian is given by the cumulant  expansion of the characteristic function $\mathcal{G}_{N}$:
	\begin{equation}\label{eq:htau}
		\mathcal{H}_{N}[\{\tau\}]=-\frac{1}{\tilde{\beta}}\log{\mathcal{G}_{N}[\{\tau\};\beta,n]} \,\,.
	\end{equation}
Here,  $\tilde{\beta}$ is the temperature parameter redefined in accordance with the corresponding redefinition of the fluid chemical potential $\mu$. The latter  plays the role of the external ``magnetic'' $\tau$-field. Certainly, such a formal  consideration cannot be considered as a proof for \eqref{eq:projtransfr},  but it demonstrates the very possibility to state the exact isomorphism between continuous fluid and discrete lattice gas or some pseudo-spin model. The linearity of the density diameter, as we have seen above, plays an important role here, simplifying the resulting transformation. Thus, the relations between thermodynamic potentials of a fluid and isomorphic lattice system suggested in \cite{eos_zenomeunified_jphyschemb2011}  can possibly be derived this way. 
\section{Conclusion}\label{sec:end}
The results presented show that the law of rectilinear diameter augmented with the global isomorphism approach can be effectively used to describe  the liquid-gas equilibrium in fluids on the basis of information for the lattice gas. Additionally, we note that the Nishimory line in spin glass systems \cite{crit_nishimori_ptp1981} is very similar to the Zeno-line. Along both lines, we have some kind of the  ``ideal-gas'' behaviour, which means that the calculation of the partition function becomes miraculously trivial. In essence, the global isomorphism approach states that the concept of the symmetrical order parameter widely used in the vicinity of the critical point can be extended to the whole region of liquid-gas equilibrium. The existence of such a canonical ``magnetization''  order parameter seems trivial if the binodal diameter is analytic in temperature. Note that the absence of particle-hole symmetry does not necessarily mean the non-analytic behaviour~\cite{crit_diamvausesak_jpmathgen1980} but rather means that density fluctuations are orthogonal to entropy ones \cite{book_patpokr}. In general, the order parameter is the composite field of density and entropy fluctuations, so the  effective Hamiltonian is a functional of these two fields. In comparison with density fluctuations, the ones of entropy  are short-ranged and weaker. Therefore, integration over them effectively leads to the asymmetry of the density field Hamiltonian, which is commonly represented by $\varphi^4$ field theory model. However, as was shown in~\cite{crit_nicoll_pra1981} (see also \cite{crit_can_newham_jmolphys2010,crit_asymbertranisnicoll_pre2012}) the inclusion of asymmetrical terms like $\varphi^5$ is also needed, which obviously requires the consideration of asymmetry within  $\varphi^6$ Hamiltonian. The latter appears naturally also in I.~R.~Yukhnovskii’s collective variable approach (see, e.g., \cite{crit_phi6collvar_prb2002}). In this sense, our consideration of the asymmetry of polymer solution equilibrium, with parameter $z$ related to the polymerization degree, gives a unified view, bearing in mind that for molecular fluids $z$ depend on their association degree \cite{eos_zenozcassocme_jcp2014}. Indeed,  the effective field theory for the tricriticality in the Flory $\theta$-point is described by  $\varphi^6$ Hamiltonian~\cite{crit_degennestricritpoly_jphyslet1975}. Thus, we hope that the global isomorphism approach provides a unified scheme for the consideration of fluid asymmetry. We are going to explore these topics in future works. 
	\section*{Acknowledgements}
	This work was supported by the fortitude of the Armed Forces of Ukraine.
	\bibliographystyle{cmpj}
	%
%

%
\newpage
	\ukrainianpart
	\title{Регулярності рідинно-газового стану як прояв глобального ізоморфізму з моделлю Ізінга}
	\author{Л. А.~Булавін\refaddr{knu}, В. Л.~Кулінський\refaddr{onu,ontu}, А. М. Катц\refaddr{onu}, А. М.~Маслечко\refaddr{onu}}
	\addresses{
		\addr{knu} Київський національний університет імені Тараса Шевченка, просп. академіка Глушкова, \\03022 Київ, Україна
		\addr{onu} Одеський національний університет імені І. І. Мечникова, вул. В.~Змієнка 2,  65026 Одеса, Україна
		\addr{ontu} Одеський національний технологічний університет, вул. Канатна 112, 65039 Одеса, Україна}
	\makeukrtitle
	\begin{abstract}
		\tolerance=3000%
		Рівновага рідина-газ розглядається в рамках глобального ізоморфізму за допомогою моделі Ізінга (ґратчастий газ). Такий ізоморфізм можна розглядати з огляду на існування параметра порядку, за яким симетрія бінодалі відновлюється не тільки в околі критичної точки (критичний ізоморфізм), але й глобально у всій області співіснування. Ми показуємо, як емпіричний закон прямолінійного діаметра густини рідинно-газової бінодалі дозволяє вивести досить просту форму перетворення ізоморфізму між рідиною та моделлю граткового газу типу Ізінга. Співвідношення для критичних параметрів, які випливають із такого ізоморфізму, перевіряються на різноманітних рідинних системах як реальних, так і модельних. Також ми розглядаємо фазову рівновагу в розчинах полімерів і  $\theta$-точку Флорі як крайній стан такої рівноваги в рамках нашого підходу. Найважливішим є тестування у випадку 2D з використанням точного розв'язку Онзагера моделі Ізінга, і ми представляємо результати нашого підходу до розрахунку параметрів критичної точки моношарів для благородних газів і поверхневого натягу двовимірного флюїду.
		\keywords глобальний ізоморфізм, рівновага рідина-пара, прямолінійний діаметр, Зено-лінія, модель Ізінга
	\end{abstract}
\end{document}